\documentclass[english]{article}
\usepackage{mathptmx}
\usepackage{helvet}
\usepackage[T1]{fontenc}
\usepackage[latin9]{inputenc}
\usepackage{geometry}
\usepackage{amsmath}
\usepackage{graphicx}
\usepackage{setspace}
\makeatletter
\newcommand{\lyxaddress}[1]{
\par {\raggedright #1
\vspace{1.4em}
\noindent\par}
}

\makeatother

\usepackage{babel}
\begin{document}

\title{Stability analysis of a viscoelastic model for ion-irradiated silicon}

\author{Scott A. Norris}

\maketitle

\lyxaddress{Department of Mathematics, Southern Methodist University, Dallas
TX 75205 USA}
\begin{abstract}
To study the effect of stress within the thin amorphous film generated
atop Si irradiated by Ar\textsuperscript{+}, we model the film as
a viscoelastic medium into which the ion beam continually injects
biaxial compressive stress. We find that at normal incidence, the
model predicts a steady compressive stress of a magnitude comparable
to experiment. However, linear stability analysis at normal incidence
reveals that this mechanism of stress generation is unconditionally
stabilizing due to a purely kinematic material flow, depending on
none of the material parameters. Thus, despite plausible conjectures
in the literature as to its potential role in pattern formation, we
conclude that beam stress at normal incidence is unlikely to be a
source of instability at any energy, supporting recent theories attributing
hexagonal ordered dots to the effects of composition. In addition,
we find that the elastic moduli appear in neither the steady film
stress nor the leading order smoothening, suggesting that the primary
effects of stress can be captured even if elasticity is neglected.
This should greatly simplify future analytical studies of highly nonplanar
surface evolution, in which the beam-injected stress is considered
to be an important effect.
\end{abstract}

\section{Introduction }

Pattern formation resulting from uniform ion irradiation of solid
surfaces represents a promising potential route to controlled nano-scale
surface modification. In particular, the low energy regime (typically
$10^{2}-10^{4}$ eV), where the energy loss is dominated by nuclear
collision cascades, has been the topic of continued experimental and
theoretical investigations. Due to its simplicity, noble-gas ion irradiation
of silicon has been extensively studied as a very promising system
for experimental tests of theory: it is a monatomic system amenable
to molecular dynamics simulation and its near-surface region is amorphous
under ion bombardment, thereby minimizing the potentially confounding
effects of disproportionation and crystallographic singularities \cite{chan-chason-JAP-2007,kalff-comsa-michely-SS-2001}. 

Despite its attractive attributes, the noble gas / silicon system
has proven remarkably finicky, confounding researchers via inter-laboratory
irreproducibilities. In particular, for normal-incidence ion-irradiation,
researchers in various groups at various times have observed either
hexagonal arrays of dots \cite{gago-etal-APL-2001}, disordered ripple
structures \cite{madi-etal-2008-PRL}, combinations of dots and ripples
\cite{george-thesis-2007}, or featureless flat surfaces \cite{madi-aziz-ASS-2011}.
The most current physical models of pure materials \cite{davidovitch-etal-PRB-2007,zhou-etal-PRB-2008,norris-etal-NCOMM-2011}
-- coupling erosion \cite{bradley-harper-JVST-1988}, mass redistribution
\cite{carter-vishnyakov-PRB-1996}, and ion-enhanced viscous flow
\cite{umbach-etal-PRL-2001} -- have been shown to be maximally stable
at normal incidence \cite{norris-etal-NCOMM-2011}, suggesting that
flat surfaces should be generically observed. This has led to speculation
that additional physical effects may be generating the observed structures,
such as long-range atomic redeposition \cite{facsko-etal-PRB-2004},
thin-film stress \cite{madi-etal-2008-PRL}, or the effect of contaminants
\cite{macko-etal-NanoTech-2010}. 

There is growing evidence that structures observed on Si under normal-incidence
ion irradiation are due to experimental contaminants. On the one hand,
it has been shown that after the careful removal of contaminants \cite{madi-etal-2008-PRL,madi-etal-JPCM-2009}
and other experimental artifacts \cite{madi-aziz-ASS-2011}, formerly
patterned surfaces become flat. On the other hand, the controlled
addition of contaminants to pure surfaces causes patterns to emerge
\cite{ozaydin-etal-APL-2005,ozaydin-etal-JVSTB-2008,macko-etal-NanoTech-2010}.
Finally, a recent model of concentration effects does admit an instability
at normal incidence \cite{bradley-shipman-PRL-2010}. These results
represent strong evidence for the impurity-driven theory of structure
formation. 

To isolate impurities as the sole cause of these structures, it is
desirable to rule out all other proposed candidates. Very recently,
Bradley \cite{bradley-PRB-2011} has shown that redeposition is a
nonlinear effect and therefore cannot contribute to linear stability.%
\footnote{In principle, there is the possibility that sputtered ions charged
with one sign or the other will be sufficiently attracted to the surface
(depending on the surrounding field lines) to redeposit at distances
large with respect to the instability wavelength. However, due to
the concentration of field lines at topographical hilltops, we anticipate
that this would be a \emph{destabilizing} effect proportional to \emph{curvature},
and so it would not lead to the height-dependent, stabilizing term
hypothesized by Facsko et al. \cite{facsko-etal-PRB-2004}and discussed
by Davidovitch et al. \cite{davidovitch-etal-PRB-2007}.%
} In this paper, we show that a very general, viscoelastic model of
the amorphous surface layer into which a normal-incidence ion beam
is continually injecting biaxial stress is morphologically stable
against topographical perturbations of all wavelengths. This stands
in agreement with the most recent experimental results for this system
\cite{madi-aziz-ASS-2011}, and provides additional support for the
concentration-dependence of observed structures. In addition, we find
that the leading-order film dynamics due to beam-injected stress are
independent of the elastic constants of the film, suggesting that
elasticity may be safely neglected to first approximation.

\begin{onehalfspace}

\section{Model}
\end{onehalfspace}

We consider a two-dimensional viscoelastic film of amorphous silicon,
irradiated at normal incidence, sitting atop a rigid crystalline substrate.
For simplicity, we neglect erosion, so as to focus purely on the effect
of stress.%
\footnote{When erosion is considered, one must adopt a moving frame of reference
that follows the eroding interface. In this frame, there is a steady
{}``background'' velocity in the vertical (normal) direction. In
addition, the stress is zero at the film/substrate interface in the
absence of density change, and exponentially approaches the steady
state found here. The net effect is that the eroding film has a steady
stress of the same form, but of a smaller and position-dependent magnitude,
than that found here. The differences are small when the time it takes
material to be advected through the amorphous layer from crystalline
substrate to free surface is long compared to the Maxwell time, given
by the ratio of the shear viscosity to the shear modulus. Under the
experimental conditions described here, the advection time is approximately
15 seconds, whereas the Maxwell time is approximately 20 milliseconds.
} We choose a co-ordinate system $\left(x,z\right)$ pinned to the
the film/substrate interface $z=0$. Hence, $x$ is the lateral co-ordinate,
and $z$ is the vertical co-ordinate, with the semi-infinite crystal
occupying $z<0$. In what follows, $\mathbf{E}$ and $\mathbf{T}$
denote the strain and stress tensors, respectively, while $\mathbf{E}_{D}$
and $\mathbf{T}_{D}$ are their deviatoric components:
\begin{align*}
\mathbf{E}_{D} & =\mathbf{E}-\frac{1}{3}\mathrm{tr}\left(\mathbf{E}\right)\mathbf{I}\\
\mathbf{T}_{D} & =\mathbf{T}-\frac{1}{3}\mathrm{tr}\left(\mathbf{T}\right)\mathbf{I}.
\end{align*}
Because we are studying infinitesimal perturbations to a stationary
film, we will employ the small-strain approximation

\begin{onehalfspace}
\begin{align*}
\frac{D\mathbf{E}}{Dt} & \approx\frac{1}{2}\left(\nabla\mathbf{v}+\nabla\mathbf{v}^{\text{T}}\right),
\end{align*}
where $\mathbf{v}=\left(u,w\right)^{T}$ is the velocity vector. 
\end{onehalfspace}

The ion irradiation imparts a stress into the film, while simultaneously
enhancing the fluidity of the film; hence, a viscoelastic constitutive
model is used. In two dimensions, a simple constitutive relation for
the film is (see \cite{trinkaus-ryazanov-PRL-1995-viscoelastic,trinkaus-NIMB-1998-viscoelastic,van-dillen-etal-PRB-2005-viscoelastic-model,otani-etal-JAP-2006}):
\begin{equation}
\frac{D}{Dt}\left[\mathbf{E}\right]=\frac{1}{2\eta}\mathbf{T}_{D}+\frac{1}{2G}\frac{D}{Dt}\left[\mathbf{T}_{D}\right]+\frac{1}{9B}\frac{D}{Dt}\left[\mathrm{tr}\left(\mathbf{T}\right)\right]\mathbf{I}+fA\left[\begin{array}{ccc}
1 & 0 & 0\\
0 & 1 & 0\\
0 & 0 & -2
\end{array}\right]\label{eqn: otani-cr}
\end{equation}
The first three terms on the right-hand side of Eqn. \eqref{eqn: otani-cr}
constitute a standard Maxwell model of viscoelasticity for a two-dimensional
material with viscosity $\eta$, shear modulus $G$, and bulk modulus
$B$. The fourth term describes the imposition of a stress-free strain
by the beam, with $f$ the ion flux, $A$ a measure of strain imparted
per ion, and the matrix describing a {}``pancake strain'' -- a pure
shear consisting of compression in the vertical direction and an equal
expansion in the lateral direction.

\begin{onehalfspace}
With stress and strain defined in terms of the velocity field in \eqref{eqn: otani-cr},
the bulk governing equations are simply Newton's second law and the
conservation of mass. Assuming that the former simplifies to Stokes
flow in a limit of low Reynolds number, we thus have in the bulk 
\begin{align}
\nabla\cdot\mathbf{T} & =0\label{eqn: mom-conservation}\\
\nabla\cdot\left(\rho\mathbf{v}\right) & =0,\label{eqn: mass-conservation}
\end{align}
where $\rho$ is the density. At the boundaries, we have
\begin{align}
\mathbf{v} & =0 & \qquad & \left(\mathrm{at}\: z=0\right)\label{eqn: no-slip}\\
v_{\mathbf{n}} & =\mathbf{v}\cdot\hat{\mathbf{n}} & \qquad & \left(\mathrm{at}\: z=h\left(x\right)\right).\label{eqn: kinematic}\\
\mathbf{T}\cdot\hat{\mathbf{n}} & =-\gamma\kappa\hat{\mathbf{n}} & \qquad & \left(\mathrm{at}\: z=h\left(x\right)\right)\label{eqn: surface-stress}
\end{align}
Here \eqref{eqn: no-slip} is the no-slip condition at the film/substrate
interface $z=0$. At the free interface $z=h\left(x\right)$, $\hat{\mathbf{n}}$
is the surface normal, the kinematic condition \eqref{eqn: kinematic}
relates the velocity $v_{\mathbf{n}}$ of the free surface, normal
to itself, to the bulk material velocity field $\mathbf{v}$. Finally,
condition \eqref{eqn: surface-stress} gives the surface stress in
terms of the surface energy $\gamma$ and surface curvature $\kappa$.
\end{onehalfspace}

\section{Analysis}

\subsection{Steady Solution}

We first look for a steady state ($\partial/\partial t\to0$) consisting
of a flat film. Using translational and reflective symmetry in $x$
and $y$, we can limit the steady velocity field $\mathbf{v}_{0}$
to the form
\begin{equation}
\mathbf{v}_{0}\left(z\right)=\left(0,\,0,\, w_{0}\left(z\right)\right)^{T}.\label{eqn: steady-velocity-form}
\end{equation}
Then, conservation of mass requires that $w_{0}\left(z\right)=0$,
and so the film is stationary, as we expect. However, the \emph{strain
and stress} associated with this steady state are not determined by
the above considerations. These can be obtained as follows. First,
from the steady version of the constitutive relation \eqref{eqn: otani-cr},
we can write the steady deviatoric stress as 
\begin{equation}
\mathbf{T}_{D,0}=-2\eta fA\left[\begin{array}{ccc}
1 & 0 & 0\\
0 & 1 & 0\\
0 & 0 & -2
\end{array}\right].
\end{equation}
Hence, the steady stress tensor is 
\begin{equation}
\mathbf{T}_{0}=-2\eta fA\left[\begin{array}{ccc}
1 & 0 & 0\\
0 & 1 & 0\\
0 & 0 & -2
\end{array}\right]+\frac{1}{3}\mathrm{tr}\left(\mathbf{T}_{0}\right)\mathbf{I},
\end{equation}
where the trace of the stress tensor (the \emph{negative pressure})
is unknown. Second, we apply the surface stress condition \eqref{eqn: surface-stress}
for a flat surface to obtain a single equation 
\begin{equation}
\mathrm{tr}\left(\mathbf{T}_{0}\right)=-12\eta fA,
\end{equation}
which solves for the steady stress. Third, the spherical part of the
constitutive relation integrates to

\begin{equation}
\mathrm{tr}\left(\mathbf{E}_{0}\right)=\frac{1}{3B}\mathrm{tr}\left(\mathbf{T}_{0}\right)=-4\frac{\eta}{B}fA.
\end{equation}
Finally, the form \eqref{eqn: steady-velocity-form} for the steady
velocity limits the steady strain tensor to the form
\begin{equation}
\mathbf{E}_{0}=\left[\begin{array}{ccc}
0 & 0 & 0\\
0 & 0 & 0\\
0 & 0 & \frac{\partial w_{0}}{\partial z}
\end{array}\right],
\end{equation}
implying that $\frac{\partial w_{0}}{\partial z}=\mathrm{tr}\left(\mathbf{E}_{0}\right)=-4\frac{\eta}{B}fA.$
Collecting all of this information, we can express the steady strain
and stress as 
\begin{equation}
\begin{aligned}\mathbf{E}_{0} & =4\frac{\eta}{B}fA\left[\begin{array}{ccc}
0 & 0 & 0\\
0 & 0 & 0\\
0 & 0 & -1
\end{array}\right]\\
\mathbf{T}_{0} & =6\eta fA\left[\begin{array}{ccc}
-1 & 0 & 0\\
0 & -1 & 0\\
0 & 0 & 0
\end{array}\right]
\end{aligned}
;\label{eqn: steady-strain-stress}
\end{equation}
hence, in the steady state the material is \emph{compressively strained
vertically} by the beam, and \emph{compressively stressed laterally}.
It is notable that the steady stress does not depend on the elastic
moduli of the film -- only the viscosity. 

For one experimental measurement, the steady state \eqref{eqn: steady-strain-stress}
exhibits reasonable agreement with experiment. For Si irradiated with
Ar\textsuperscript{+} at 250 eV with a flux of $f=3.5\times10^{15}\;\text{ions /\ensuremath{\left(\text{cm}^{2}\:\text{\ensuremath{\sec}}\right)}}$,
a steady stress of 1.4 GPa is observed \cite{madi-thesis-2011}. At
this energy and flux, we have previously estimated $\eta\approx6.2\times10^{8}\;\text{Pa sec}$
(see supplement of \cite{norris-etal-NCOMM-2011}), and although measurements
of $A$ are rare, at 3 keV there has been an estimate of $A\approx5\times10^{-17}\;\text{cm}^{2}/\text{ion}$
\cite{george-etal-JAP-2010}. For these values, we obtain a prediction
of $\mathbf{T}_{0,xx}=\mathbf{T}_{0,yy}\approx0.65\;\text{GPa}$,
which is within about a factor of 2 of the observed value.

\subsection{Linear Stability}

\begin{onehalfspace}
We now study the linear stability of this system under a small perturbation
to the film/vapor interface by an infinitesimal \emph{normal mode}
in the $x$-direction:
\begin{equation}
h\left(x\right)=h_{0}+\varepsilon\exp\left(ikx+\sigma t\right)
\end{equation}
We consider the \emph{plane strain} limit of the governing equations,
and assume that in the linear regime, the strain and velocity fields
will share the same sinusoidal dependence on $x$ and $t$; we therefore
write

\begin{equation}
\left[\begin{array}{c}
\mathbf{v}_{\phantom{0}}\\
\mathbf{E}_{\phantom{0}}
\end{array}\right]=\left[\begin{array}{c}
\mathbf{v}_{0}\\
\mathbf{E}_{0}
\end{array}\right]+\varepsilon\left[\begin{array}{c}
\tilde{\mathbf{v}}\left(z\right)\\
\tilde{\mathbf{E}}\left(z\right)
\end{array}\right]e^{ikx+\sigma t}.\label{eqn: fourier-ansatz}
\end{equation}
Upon inserting the ansatz \eqref{eqn: fourier-ansatz} into the governing
equations \eqref{eqn: otani-cr}-\eqref{eqn: mass-conservation},
and keeping only terms to leading order in the infinitesimal parameter
$\varepsilon,$ we find that the perturbation $\tilde{\mathbf{v}}=\left(\tilde{u},\,\tilde{w}\right)^{T}$
to the velocity field is governed by the pair of ordinary differential
equations 
\begin{equation}
\begin{aligned}\tilde{u}''-N\tilde{w}'-K\tilde{u} & =0\\
\tilde{w}''-M\tilde{u}'-L\tilde{w} & =0
\end{aligned}
;\label{eqn: odes-for-velocity}
\end{equation}
where 
\begin{equation}
\begin{aligned}K & =\frac{4\alpha+6\beta}{3\alpha}k^{2}\\
L & =\frac{3\alpha}{4\alpha+6\beta}k^{2}
\end{aligned}
\qquad\begin{aligned}M & =-i\frac{\alpha+6\beta}{4\alpha+6\beta}k\\
N & =-i\frac{\alpha+6\beta}{3\alpha}k
\end{aligned}
\end{equation}
and 
\begin{equation}
\alpha=\frac{2\eta}{1+\frac{\eta\sigma}{G}}\qquad\beta=\frac{B}{\sigma}.
\end{equation}
These equations can be re-written as a linear system and solved using
eigenvalue analysis; the general solution can be expressed as 
\begin{equation}
\begin{aligned}\left[\begin{array}{c}
\tilde{u}\\
\tilde{w}
\end{array}\right] & =\left[\begin{array}{c}
a\\
c
\end{array}\right]\cosh\left(kz\right)+\left[\begin{array}{c}
b\\
d
\end{array}\right]\sinh\left(kz\right)\\
 & +\frac{\alpha+6\beta}{7\alpha+6\beta}kz\left[\begin{array}{c}
\left(b-ic\right)\cosh\left(kz\right)+\left(a-id\right)\sinh\left(kz\right)\\
\left(-ia-d\right)\cosh\left(kz\right)+\left(-ib-c\right)\sinh\left(kz\right)
\end{array}\right]
\end{aligned}
,\label{eqn: ode-general-solution}
\end{equation}

\end{onehalfspace}

To obtain the four unknowns $\left\{ a,\, b,\, c,\, d\right\} $,
we must apply the linearized boundary conditions. From Eqn.~\eqref{eqn: no-slip}
at $z=0$, we immediately find that $a=c=0$. Turning next to Eqn.~\eqref{eqn: surface-stress}
at $z=h$, we find that its linearization is 
\begin{equation}
\begin{aligned}\tilde{T}_{xz} & =\frac{\alpha}{2}\left(ik\tilde{w}+\tilde{u}'\right) & =-6fA\eta ik\\
\tilde{T}_{zz} & =\frac{\alpha}{3}\left(-ik\tilde{u}+2\tilde{w}'\right)+\beta\left(ik\tilde{u}+\tilde{w}'\right) & =-\gamma k^{2}
\end{aligned}
.\label{eqn: linearized-free-boundary}
\end{equation}
Because $a=c=0$, Eqn.~\eqref{eqn: linearized-free-boundary} represents
a matrix equation for $b$ and $d$; solution of this equation yields
\begin{equation}
\begin{aligned}b & =-\frac{iak}{\Delta}\left\{ 6fA\eta k\left[V\cosh\left(Q\right)-UQ\sinh\left(Q\right)\right]+\gamma k^{2}\left[-\left(V-U\right)\sinh\left(Q\right)+UQ\cosh\left(Q\right)\right]\right\} \\
d & =-\frac{ak}{\Delta}\left\{ 6fA\eta k\left[-\left(V-U\right)\sinh\left(Q\right)-UQ\cosh\left(Q\right)\right]+\gamma k^{2}\left[V\cosh\left(Q\right)+UQ\sinh\left(Q\right)\right]\right\} 
\end{aligned}
,\label{eqn: ode-coefficients}
\end{equation}
where
\[
\begin{aligned}Q & =kh\\
U & =\frac{\alpha+6\beta}{7\alpha+6\beta}\\
V & =\frac{4\alpha+6\beta}{7\alpha+6\beta}
\end{aligned}
\]
are common dimensionless groups, and
\begin{equation}
\Delta=\left(\alpha k\right)^{2}\left[V^{2}+U\sinh^{2}\left(Q\right)+U^{2}Q^{2}\right]
\end{equation}
is the determinant of the matrix associated with equation \eqref{eqn: linearized-free-boundary}.
Finally, inserting the coefficients \eqref{eqn: ode-coefficients}
into \eqref{eqn: ode-general-solution}, we apply the linearized version
of the kinematic condition \eqref{eqn: kinematic},
\begin{equation}
\sigma=\tilde{w}\left(h\right),
\end{equation}
which provides the implicit dispersion relation between the growth
rate $\sigma$ and wavenumber $k$:
\begin{equation}
\begin{aligned}\frac{2R}{1+R}\left[V^{2}+U^{2}Q^{2}+U\sinh^{2}\left(Q\right)\right]\\
+D\left[U^{2}Q^{2}-\left(V-U\right)\sinh^{2}\left(Q\right)\right]\\
+CVQ\left[\sinh\left(2Q\right)-2UQ\right] & =0
\end{aligned}
.\label{eqn: dispersion-relation}
\end{equation}
Here we have converted to the dimensionless parameters $\left\{ R,\, Q,\, D,\, C\right\} $
are given by 
\begin{equation}
\begin{aligned}R & =\frac{\eta}{G}\sigma &  & \text{(growth rate)}\\
Q & =hk &  & \text{(wavenumber)}\\
D & =\frac{6fA\eta}{G} &  & \text{(Deborah number)}\\
C & =\frac{\gamma}{2Gh} &  & \text{(Capillary number)}
\end{aligned}
\label{eqn: dimensionless-parameters}
\end{equation}

\subsection{Interpretation}

Equation \eqref{eqn: dispersion-relation} is our central theoretical
result, but requires some further examination. Although an explicit
dispersion relation is not available, we can perform neutral stability
analysis on \eqref{eqn: dispersion-relation} by setting $\sigma\to0$
and solving the resulting expression for $D$, which value of $D$
we name $D^{*}$. In this limit one can show that $U\to1$, $V\to1$,
$R\to1$, and the resulting expression for $D^{*}$ is
\begin{equation}
D^{*}\left(Q\right)=2C\left[1-\frac{\sinh\left(2Q\right)}{2Q}\right];\label{eqn: neutral-stability}
\end{equation}
this value of $D$ establishes the \emph{neutral stability boundary.}
For values of $D^{*}$ in the \emph{domain} of this function, both
stable and unstable wavenumbers $Q$ exist; hence, the extremal values
of $D$ serve as boundaries between stable and unstable regions of
parameter space. As observed in Figure 4, $D^{*}\left(Q\right)$ is
a strictly negative function of $Q$, with a global maximum of $D=0$
at $Q=0$, and so the stability of the film depends upon the sign
of $D$. By implicitly differentiating \eqref{eqn: dispersion-relation}
in $R$ and $D$, we find that $\frac{\partial R}{\partial D}$ is
negative at $R=D=Q=0$, so that positive $D$ implies negative $R$.
Because $D$ is a physical constant and positive by definition, we
conclude that the film is stable at all wavelengths. Hence, even though
the pancake strain places the film in a state of compressive stress,
the film is \emph{unconditionally stable to perturbations}. Our stability
result may be understood intuitively by noting that even though the
beam stresses the bulk material below the valleys of a perturbation,
the effect on the hilltops of a small perturbation is for them to
shorten and widen under the stress-free pancake strain.

\begin{figure}
\begin{centering}
\includegraphics[width=4in]{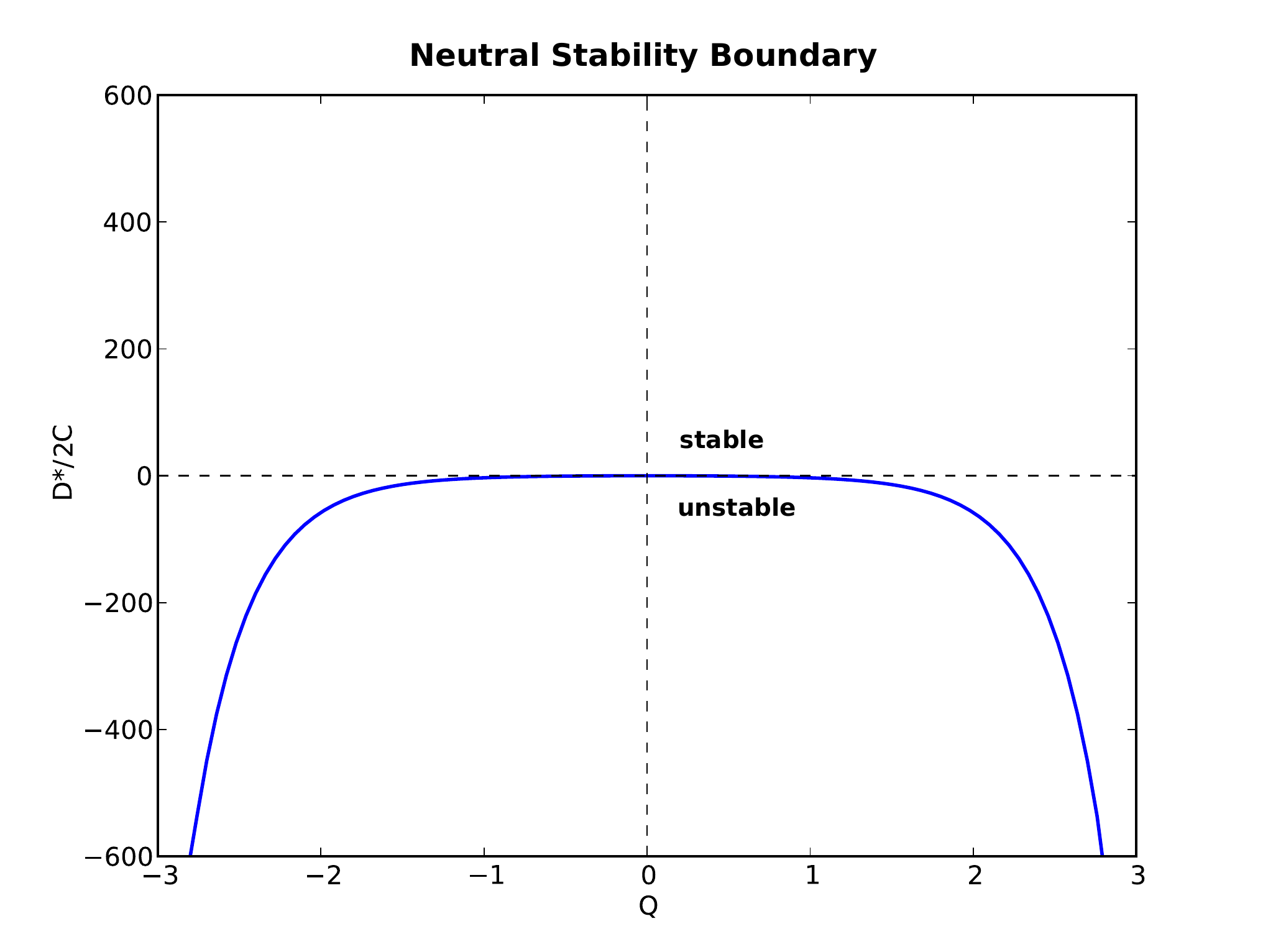}
\par\end{centering}

\label{fig: neutral-stability}\caption{Neutral Stability curve $D^{*}\left(Q\right)$, from Eqn. \eqref{eqn: neutral-stability}.}
\end{figure}

Further quantitative understanding is available for the commonly-observed
situation in which the film thickness is much smaller than the perturbation
wavelength -- i.e., that $Q=hk\ll1$. In the limit of long wavelengths
and slow evolution ( $Q\ll1$ and $R\ll1$,), the dispersion relation
\eqref{eqn: dispersion-relation} reduces, keeping the lowest order
of $Q$ in each of the coefficients, to 
\begin{equation}
R\approx-\frac{1}{2}D\, Q^{2}-\frac{2}{3}C\, Q^{4}
\end{equation}
or, reverting to dimensional form, 
\begin{equation}
\sigma=-3fA\left(hk\right)^{2}-\frac{\gamma}{3\eta h}\left(hk\right)^{4}.
\end{equation}
Hence, for common case of long-wavelength perturbations, the leading-order
contribution of the pancake strain at normal incidence is a second-order
smoothing of perturbations. A very important property of this smoothing
is that it depends on \emph{none of the bulk material properties}
of the film -- it is a purely kinematic response to the biaxial stress
injected by the beam. 

Because neither the steady stress, nor the leading order stability
properties of the film depend on the elastic properties of the medium,
it is reasonable to consider neglecting elasticity altogether. Hence,
we conclude by considering the limit of a film that is purely viscous
and incompressible ($G\to\infty$ and $B\to\infty$).%
\footnote{Two comments are in order here. First, although these limits appear
problematic in the dimensionless parameters \eqref{eqn: dimensionless-parameters},
they are in fact well-defined if the dispersion relation \eqref{eqn: dispersion-relation}
is first converted to dimensional form. Second, while the incompressible
limit is a singular limit at the level of the governing equations,
requiring the introduction of a hydrodynamic pressure, it is an ordinary
limit at the level of the solution to those equations. Hence, a re-analysis
using the incompressible equations is unnecessary.%
} In that limit, we recover the (dimensional) result 
\begin{equation}
\sigma\left(k\right)=-\frac{6fA\left(hk\right)^{2}+\frac{k\gamma}{2\eta}\left(\sinh\left(2hk\right)-2hk\right)}{\left(1+2\left(hk\right)^{2}+\cosh\left(2hk\right)\right)}.
\end{equation}
This is again unconditionally stable, and in the absence of the beam
($f\to0$), it reduces to the classic result of Orchard for viscous
surface leveling on films of arbitrary thickness \cite{orchard-ASR-1962}.

\section{Summary}

As a model for amorphous ion-irradiated solids, we have studied the
dynamics of a thin viscoelastic film subject to continual injection
of biaxial stress, and obtained two primary results. 
\begin{itemize}
\item First, we have shown that biaxial compressive stress injected into
an amorphous film by the ion beam is unconditionally stabilizing at
normal incidence, and hence that smooth surfaces at low angles should
be generic for pure amorphous materials under energetic particle irradiation.
Together with growing experimental consensus that normal-incidence
patterns only appear when contaminants are present, and Bradley's
recent demonstration that a simple model of concentration effects
does admit an instability at normal incidence \cite{bradley-shipman-PRL-2010},
this strengthens the case that these structures are due entirely to
concentration effects.
\item Second, we have shown that the leading-order contributions to film
dynamics in the small-curvature limit of this model are independent
of the elastic constants. Rather than being due to elasticity, the
steady stress observed in this model is due to viscous resistance
to the stress-free strain induced by the beam. This result suggests
that elasticity may be safely neglected in future analytical efforts.
In particular, as demonstrated by George et al. \cite{george-etal-JAP-2010}
a purely viscous version of Equation \eqref{eqn: otani-cr} will greatly
simplify the analysis of highly-nonplanar surface evolution.
\end{itemize}
Although our analysis is restricted to one independent spatial dimension
and does not account for the advection in a moving reference frame
due to sputter erosion, we anticipate that the conclusions drawn here
will be no different from a deeper analysis that accounts for these
effects.

\section{Acknowledgments}

S.A.N. was partially supported by the U.S. National Science Foundation
MSPRF program, under award \#0802987. The author thanks Michael J.
Aziz, Michael P. Brenner, Ken Kamrin, and Charbel Madi for helpful
discussions.

\begin{onehalfspace}

\bibliographystyle{unsrt}
\bibliography{/home/snorris/Dropbox/research/bibliography/tagged-bibliography}
\end{onehalfspace}

\end{document}